\documentstyle[proceedings,epsf]{crckapb} 

\makeatletter                                            %KT
\@ifundefined{epsfbox}{\@input{epsf.sty}}{\relax}        %KT
\def\plotone#1{\centering \leavevmode                    %KT
\epsfxsize=\columnwidth \epsfbox{#1}}                    %KT
\def\plotone_reduction#1#2{\centering \leavevmode        %KT
\epsfxsize=#2\columnwidth \epsfbox{#1}}                  %KT
\makeatother                                             %KT
\begin{opening}
\title{SPIRAL STRUCTURE IN IP PEG:\protect\\
       CONFRONTING THEORY AND OBSERVATIONS}
\author{Takuya Matsuda}
\author{Makoto Makita}
\author{Hiroshi Yukawa}
\institute{Department of Earth and Planetary Sciences, Kobe University, Kobe 
657-8501, Japan}
\author{Henri M.J. Boffin}
\institute{Department of Physics and Astronomy, University of Wales, Cardiff 
CF2 3YB, UK}
\end{opening}
\runningtitle{Spiral Structure in IP Peg}

\begin{document}
\vspace{-0.5cm}
\begin{abstract}
The first convincing piece of evidence of spiral structure in the accretion 
disc in IP Pegasi was found by Steeghs et al. (1997). We performed two kinds 
of 2D hydrodynamic simulations, a SFS finite volume scheme and a SPH scheme, 
with a mass ratio of 0.5. Both results agreed well with each other. We 
constructed Doppler maps and line flux-binary phase relations based on density 
distributions, the results agreeing well with those obtained by observation.
\end{abstract}

\section{Introduction}

The standard model of accretion discs is the $\alpha$ disc model proposed by 
Shakura and Sunyaev (1973). One of various alternative models is the spiral 
shock model, which was first proposed by one of the present authors (Sawada, 
Matsuda \& Hachisu, 1986a, b; Sawada et al., 1987). The $\alpha$ disc model 
essentially predicts the axi-symmetric structure of the disc except for the 
stream from the L1 point, while the spiral shock model predicts a bi-symmetric 
structure except for the stream. This fact is very important in order to 
distinguish between the two models observationally.

Steeghs, Harlaftis \& Horne (1997) found the first convincing piece of evidence 
of spiral structure in the accretion disc of the eclipsing dwarf nova binary 
IP Pegasi using the technique known as Doppler tomography. IP Pegasi consists 
of a 1.02 $M_{\odot}$ white dwarf and a 0.5 $M_{\odot}$ companion star. 

\section{Model and methods of calculation}

We calculated two-dimensional flows in a compact binary system by two numerical 
methods: the Simplified Flux vector Splitting (SFS) scheme (Jyounouchi et al., 
1993; Shima \& Jyounouchi, 1994) and the SPH scheme used by Yukawa, Boffin 
and Matsuda (1997). We studied a case with a mass ratio equal to 0.5 in order 
to simulate the IP Peg.

Only the region surrounding the white dwarf was calculated. The origin of the 
coordinates was at the center of the white dwarf, and the computational region 
was $-0.57\leq x, y \leq 0.57$. In the case of our SFS scheme, the region was 
divided into $228 \times 228$ grid points. On the other hand, the number of 
particles in the SPH scheme was about 20,000. We assumed an ideal gas which 
was characterized by a ratio of specific heats $\gamma$. The $\gamma$ was assumed 
to be 1.2 in the present calculations. The gas was injected from a small hole 
at the L1 point, which was at $x=-0.57, y=0$. After a few orbital periods, 
we obtained a nearly steady density pattern.

\section{Comparison with the observation}

Based on the density distributions obtained by the SFS and SPH schemes, we 
constructed two Doppler maps, which showed the density distributions in the 
velocity space. 
Steeghs et al. (1997) observed the line intensities of H$_\alpha$ and HeI. 
Therefore, strictly speaking, we 
had to calculate the line intensity distributions based on the temperature 
distribution. Since our calculations were two-dimensional, our temperature 
distribution might not be very relevant to construct realistic Doppler maps. 
Therefore, we used two-dimensional density distributions instead. Line 
flux-binary phase diagrams could also be produced from the calculated Doppler 
maps.

Figure 1 shows the observed line flux as a function of a binary phase with 
H$_\alpha$ and HeI obtained by Steeghs et al. (1997). Figure 2 shows our 
calculated line flux-binary phase relations (top row), calculated Doppler maps 
(middle row), and calculated density distribution (bottom row). We used two 
methods of calculation. The left column of Fig. 2 shows the results based on 
the SFS scheme, while the right column shows those obtained by the SPH scheme.
In constructing calculated Doppler maps, a contribution was added due to the 
companion star. The contribution from the outer region of the accretion disc, 
$r>0.22$, was omitted in order to avoid the effect of the stream from the L1 
point. This contribution is not seen in observation, a point which has not 
been explained yet. As can be seen, the calculated line flux and the Doppler 
maps shown in Fig. 2 agree very well with observations found in Fig. 1.

\section{Discussion}

Godon, Livio, and Lubow (1998) recently carried out a work similar to ours, 
although they only showed the line of peak density in the Doppler map. They 
claim that the calculated disc, which best fits the observation, is very hot 
compared with that in other observations. 
This claim is true in our calculations as well, in the sense that $\gamma=1.2$ produces a rather hot disc. We performed 
3D calculations (see Makita and Matsuda in this volume), in which we obtained 
rather open spirals even for the case of $\gamma=1.01$, which gives a cooler 
disc. Therefore, it might be speculated that the 3D disc is more promising 
than the 2D disc as a means to explain the observations.

\vspace{-8mm}
\begin{figure}[hbt]
\begin{center}
\leavevmode
\epsfxsize=80mm
\epsfbox{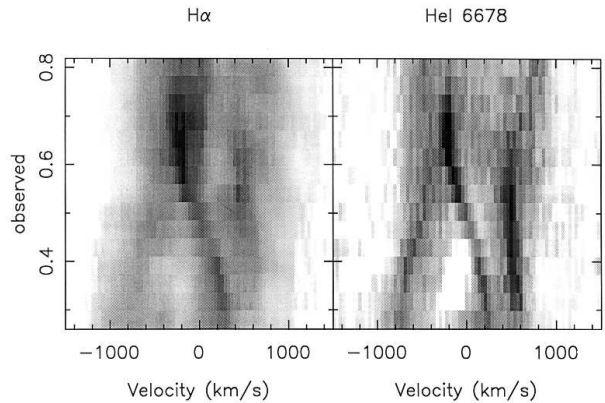}
\end{center}
\caption{The observed line flux from IP Peg as a function of the binary phase 
with H$_\alpha$ on the left and HeI(6678) on the right; these figures are taken 
from Steeghs et al.(1997) }
\end{figure}

\vspace*{-30mm}
\begin{figure}[hbt]
\begin{center}
\leavevmode
\epsfxsize=50mm
\epsfbox{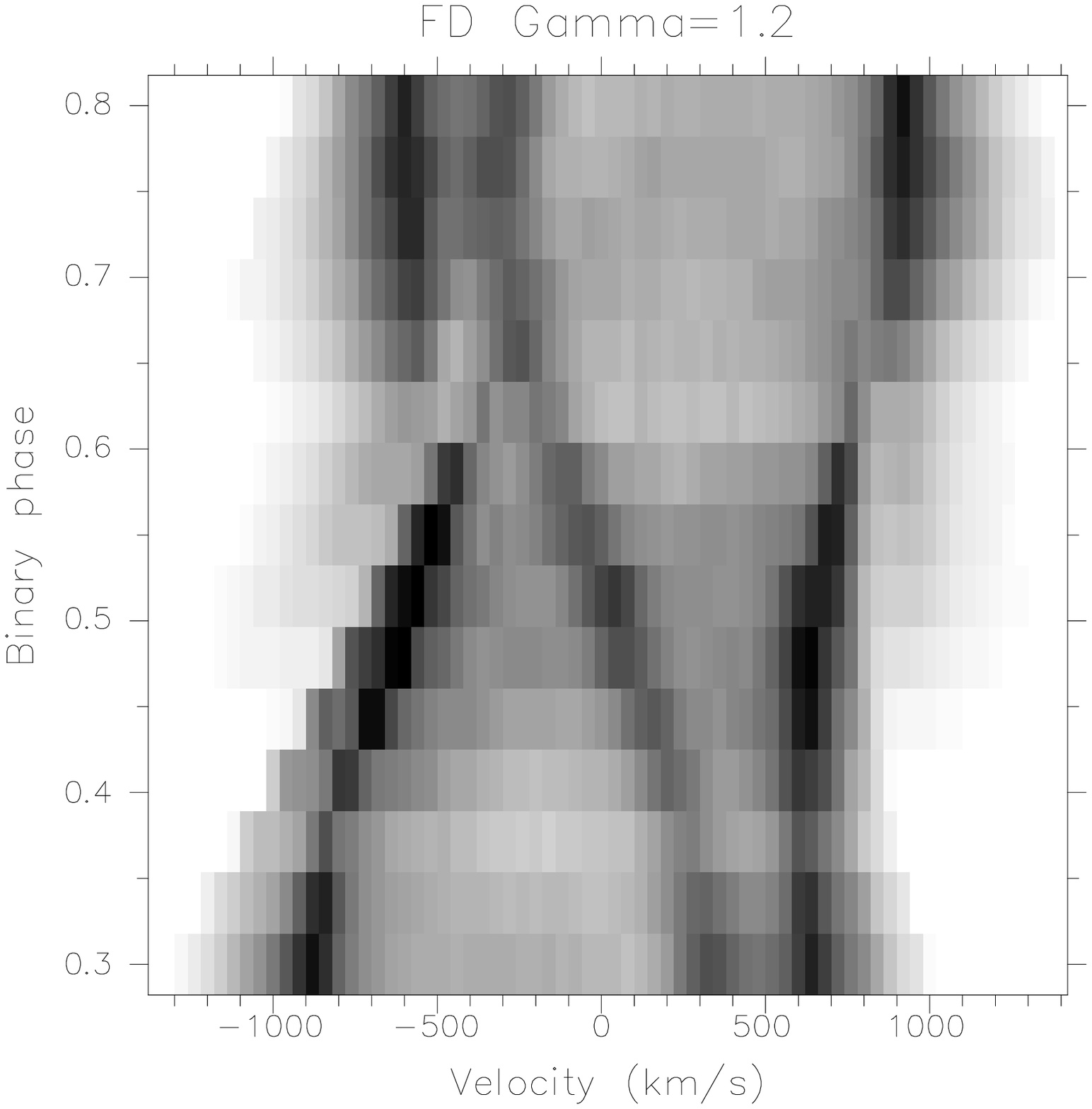}
\epsfxsize=50mm
\epsfbox{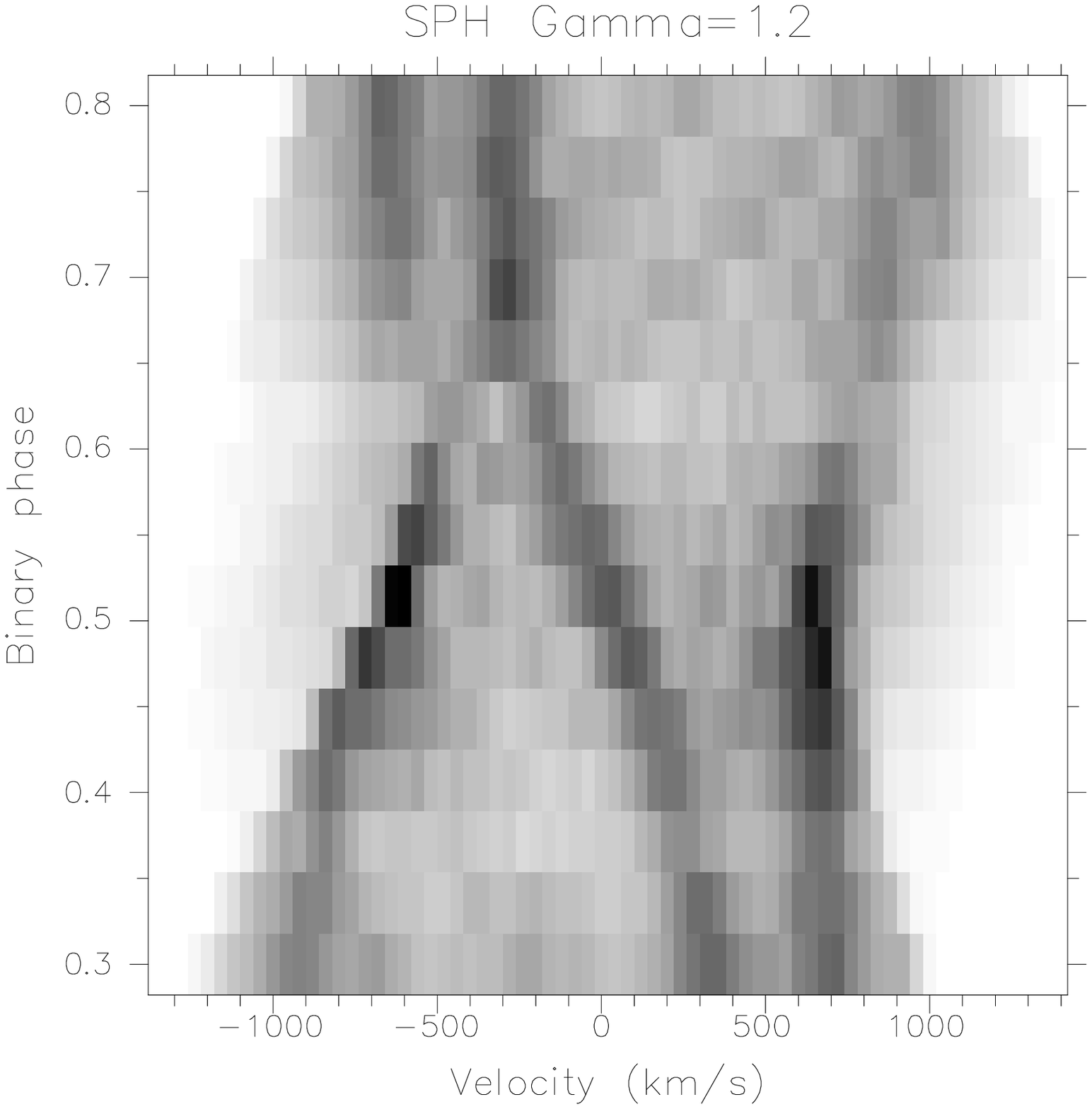}
\vspace{-3mm}
\epsfxsize=50mm
\epsfbox{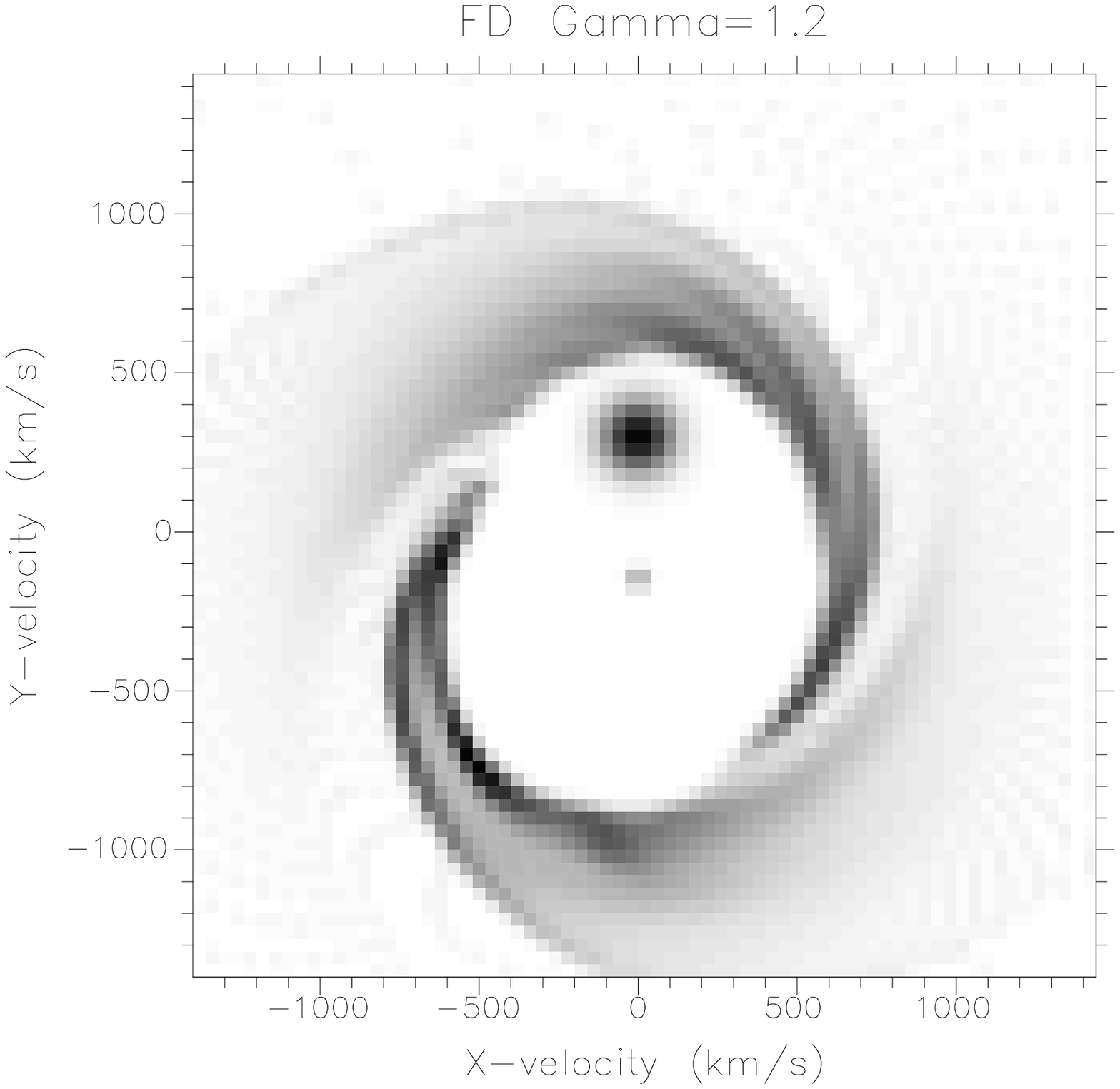}
\epsfxsize=50mm
\epsfbox{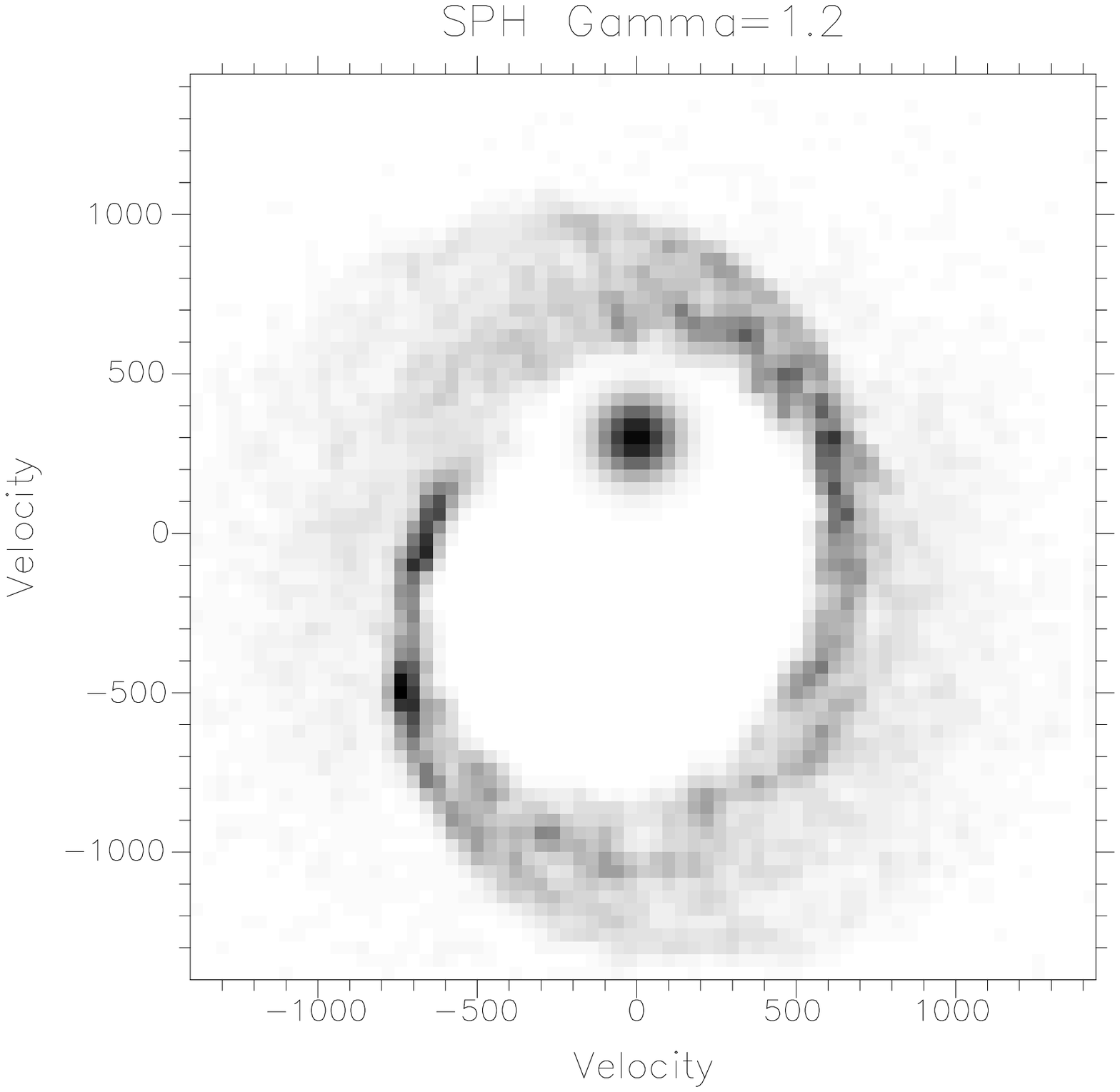}
\vspace{-3mm}
\hspace*{2.2mm}
\epsfxsize=45mm
\epsfbox{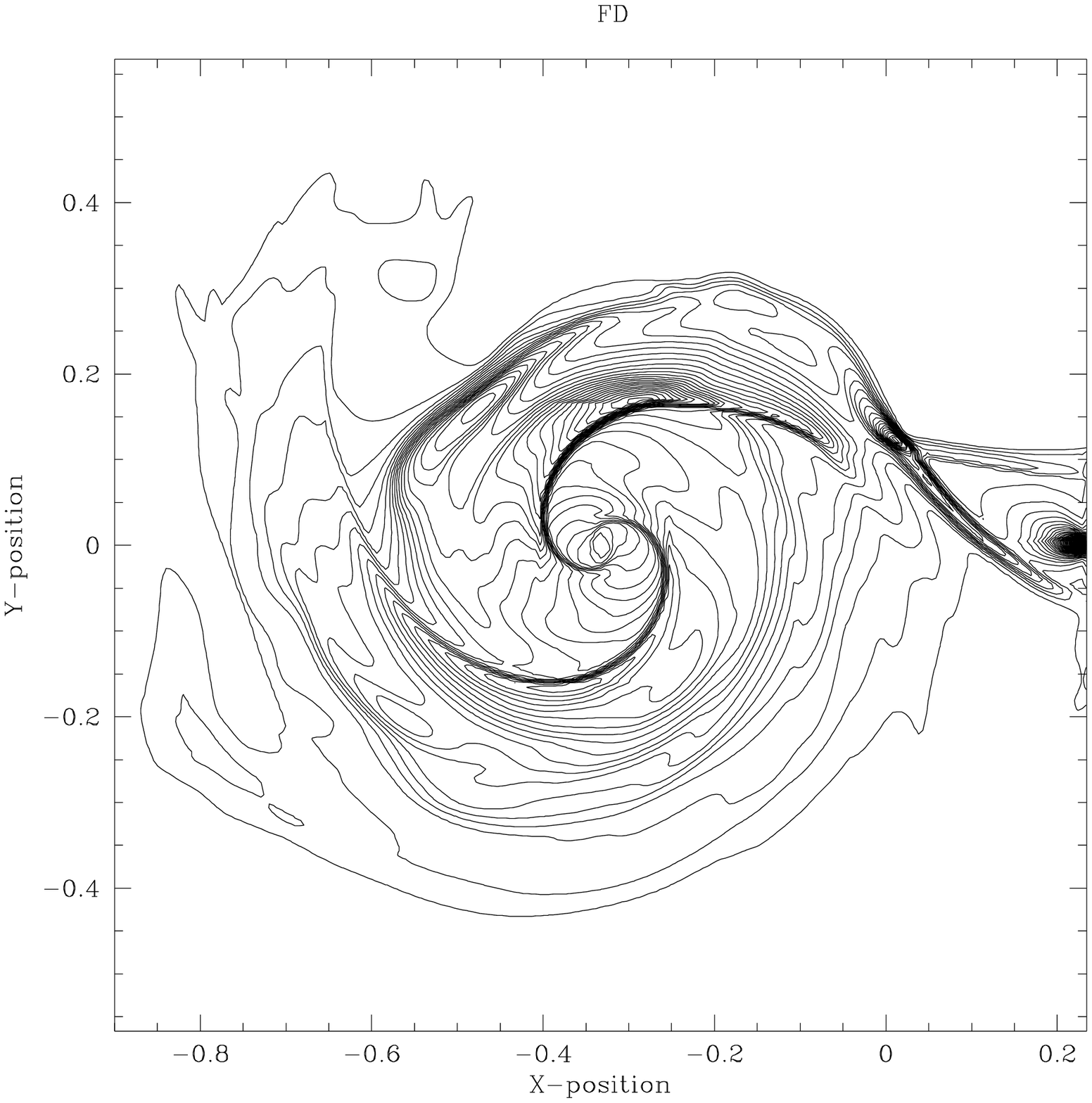}
\hspace{5mm}
\epsfxsize=45mm
\epsfbox{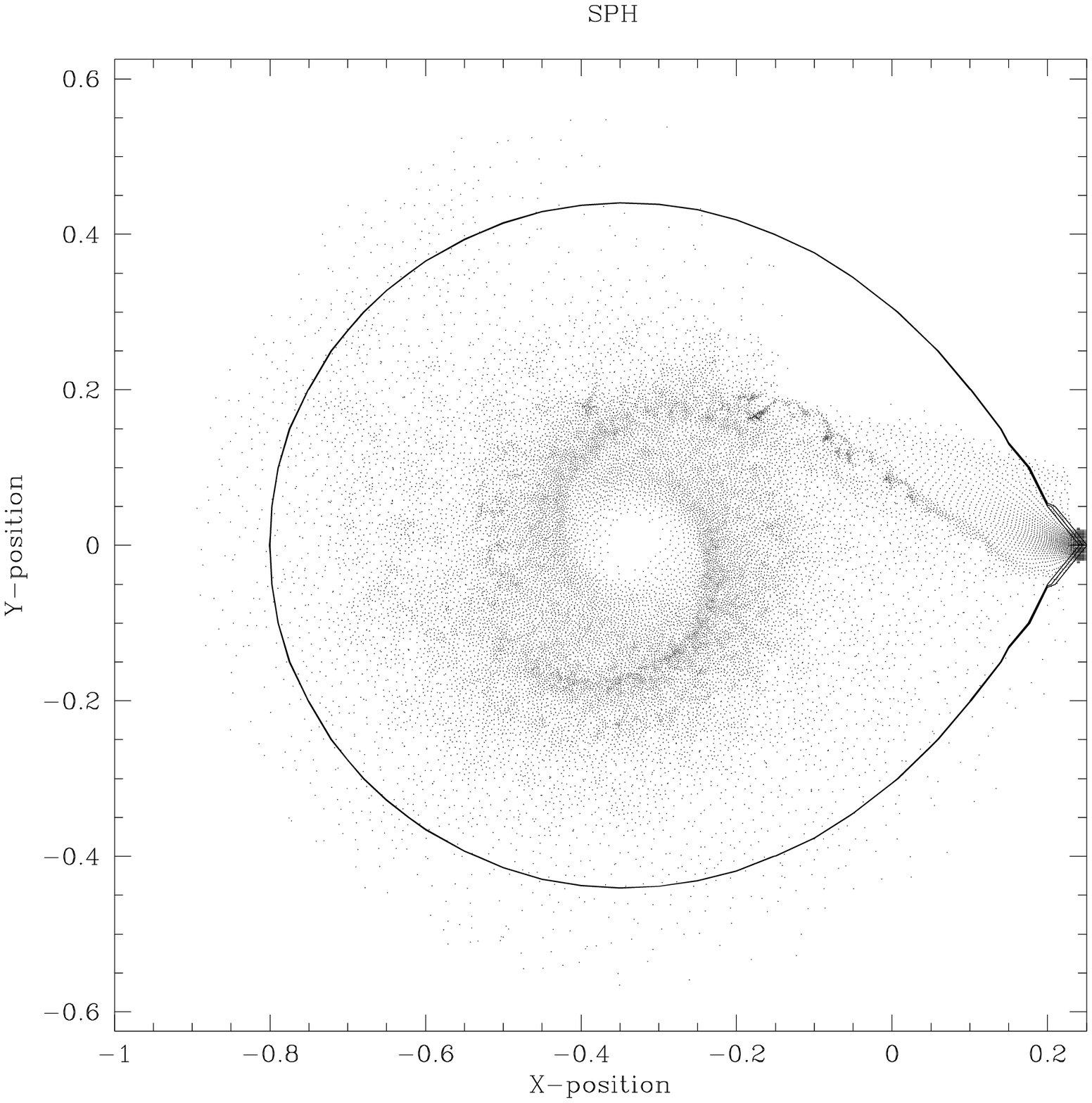}
\end{center}

\caption{Top row: calculated line flux as a function of the binary phase 
(vertical axis). The horizontal axis represents the radial velocity. Middle 
row: calculated Doppler map. The horizontal axis and the vertical axis depict 
the horizontal velocity, $v_x$, and the vertical velocity, $v_y$, respectively. 
The bottom row shows the density distribution. The left column shows the results 
based on the SFS method, while the right column shows those based on the SPH 
method}
\end{figure}
\end{document}